\documentclass[preprint]{aastex}
\slugcomment{To appear in the Astrophysical Journal}

\shorttitle{Infall to the Keplerian Disk in L1551 NE}
\shortauthors{Takakuwa et al.}

\begin{document}

%% LaTeX will automatically break titles if they run longer than 
%% one line. However, you may use \\ to force a line break if 
%% you desire.

\title{Evidence for Infalling Gas of Low Angular Momentum\\
towards the L1551 NE Keplerian Circumbinary Disk}
\author{Shigehisa Takakuwa\altaffilmark{1}, Masao Saito\altaffilmark{2}, Jeremy Lim\altaffilmark{3},
\& Kazuya Saigo\altaffilmark{4}}
\altaffiltext{1}{Academia Sinica Institute of Astronomy and Astrophysics, P.O. Box 23-141, Taipei 10617, Taiwan;
takakuwa@asiaa.sinica.edu.tw}
\altaffiltext{2}{Joint ALMA Observatory, Ave. Alonso de Cordova 3107, Vitacura, Santiago, Chile}
\altaffiltext{3}{Department of Physics, University of Hong Kong, Pokfulam Road, Hong Kong}
\altaffiltext{4}{ALMA Project Office, National Astronomical Observatory of Japan, Osawa 2-21-1,
Mitaka, Tokyo 181-8588, Japan}

\begin{abstract}
We report follow-up observations of the Class I binary protostellar system L1551 NE in the C$^{18}$O (3--2) line with the SMA in its compact and subcompact configurations.  Our previous observations at a higher angular resolution in the extended configuration revealed a circumbinary disk exhibiting Keplerian motion.  The combined data having more extensive spatial coverage ($\sim$140--2000 AU) verify the presence of a Keplerian circumbinary disk,
%with an outer radius of $\sim$300 AU, 
and reveals for the first time a distinct low-velocity ($\lesssim$$\pm$0.5 km s$^{-1}$ from the systemic velocity) component that displays a velocity gradient along the minor axis of the circumbinary disk.  Our simple model that reproduces the main features seen in the Position-Velocity diagrams comprises a circumbinary disk exhibiting Keplerian motion out to a radius of $\sim$300~AU, beyond which the gas exhibits pure infall at a constant velocity of $\sim$0.6 km s$^{-1}$.  The latter is significantly smaller than the expected free-fall velocity of $\sim$2.2 km s$^{-1}$ onto the L1551 NE protostellar mass of $\sim$0.8 $M_{\odot}$ at $\sim$300~AU, suggesting that the infalling gas is decelerated as it moves into regions of high gas pressure in the circumbinary disk.  The discontinuity in angular momenta between the outer infalling gas and inner Keplerian circumbinary disk implies an abrupt transition in the effectiveness at which magnetic braking is able to transfer angular momentum outwards, a result perhaps of the different plasma $\beta$'s and ionization fractions between the outer and inner regions of the circumbinary disk.
%We suggest that the combination of a relatively high plasma $\beta$ and low ionization fraction makes control of the molecular gas by magnetic fields ineffective within the Keplerian radius of the circumbinary disk, whereas the opposite applies beyond the Keplerian radius.  The transition between good and poor magnetic control of the molecular gas may occur quite abruptly not only at a certain radius, but also at a particular stage in the evolution of a protostellar system as the mass and hence density of its circumbinary disk grows.  

%We estimate a mass-accretion rate onto the Keplerian circumbinary disk of $\sim$9.6 $\times$ 10$^{-7}$ $M_{\odot}$ yr$^{-1}$.  Given that the mass in the surrounding dense condensation is $\sim$0.39 $M_{\odot}$, there is sufficient material to feed the Keplerian circumbinary disk at the abovementioned mass-accretion rate for the next $\sim$4.1 $\times$ 10$^{5}$ yr, which is comparable to the typical timescale estimated for the Class I evolutionary stage.
\end{abstract}

\section{Introduction}

% Circumstellar disks are ubiquitous around young stellar objects \citep{bec90,bec96,bra00}.
Compact and flattened circumstellar structures are commonly seen around T~Tauri stars at the Class~II stage of evolution for low-mass young stellar objects.  Observations in molecular lines reveal that these structures exhibit Keplerian motion, and therefore correspond to centrifugally-supported circumstellar disks \citep{gui98,gui99,sim00,pie03,ram06}.  Observations in dust continuum emission have unveiled the presence of compact ($\lesssim$600 AU) and flattened structures also around protostars in both the Class~0 and more evolved Class~I stages \citep{loo00,bri07,eno09, mau10,eno11}.  In the case of Class~0 protostars, observations in molecular lines show primarily infalling motion only in their circumstellar disk-like structures \citep{tak07,yen10}.
When rotation also is detected, the rotational velocity is too small for the circumstellar disk-like structures to be centrifugally supported \citep{bri09,yen11,yen13}.
By contrast, an increasing number of studies have been successful at finding Keplerian motion in the circumstellar disks of Class~I objects
\citep{bri07,lom08,jor09,tob12,tak12,yen13,har13}.  The masses of their circumstellar disks span the range $\sim$0.004--0.06~$M_{\odot}$, and are therefore comparable to the masses of circumstellar disks around Class~II objects.  Their radii span the range $\sim$100--300 AU, also comparable to the sizes of circumstellar disks around Class~II objects.  Recently, Tobin et al. (2012) reported the detection of a Keplerian disk around L1527~IRS, which they claim to be an object in transition from Class 0 to I and therefore the youngest protostar known to exhibit a Keplerian circumstellar disk.   The abovementioned results suggest that the transition from primarily infalling disk-like structures to Keplerian circumstellar disks around protostars occurs sometime during their evolution from Class 0 to Class I.

%%circumstellar disks around Class II (T-Tauri) sources exhibit Keplerian motion \citep{gui98,gui99,sim00,pie03,ram06}, such observations have revealed only infalling gas around Class 0 protostars exhibiting little if any detectable rotation \citep{tak07,bri09,yen10,yen11}.
% These studies suggest that the transition from compact flattened structures around protostars to Keplerian
% circumstellar disks around Class II sources occurs sometime between the Class 0 and II stages.
%%Around Class I protostars, recent high-resolution interferometric observations in molecular lines have been increasingly successful at finding Keplerian disks \citep{bri07,lom08,jor09,tob12,tak12,yen13}. Tobin et al. (2012) have reported detection of a Keplerian disk around L1527 IRS, a transitional object from the Class 0 to I stage, and claimed that L1527 IRS is the youngest protostar known to be surrounded by a Keplerian disk.

Although theoretical models have no problem forming disk-like structures around protostars, in many models these structures are not centrifugally supported and hence do not exhibit Keplerian motion (these disk-like structures are therefore commonly referred to as pseudodisks).  Naively, if the infalling material from the surrounding dense core conserves angular momentum, eventually the rotational velocity of this material should match the Keplerian velocity and hence a centrifugally-supported disk form around the protostar.  In many theoretical models, however, the inclusion of magnetic fields so effectively transports angular momentum outwards through magnetic braking that no centrifugally-supported structures ever form (e.g., Mellon \& Li 2008; 2009; Li, Krasnopolsky, \& Shang 2011).
% \citep[e.g.,][]{mel08,mel09,zli11}.
From a series of three-dimensional MHD simulations of collapsing Bonnor-Ebert spheres, Machida et al. (2011a; 2011b) were able to successfully produce centrifugally-supported disks
around protostars.  The disks produced in their simulations, however, do not become centrifugally-supported over radial sizes of $\gtrsim$100~AU until nearly all of the material in their parental dense core has been exhausted, and furthermore have masses of a few to a few hundred times the mass of their central protostars and hence do not exhibit Keplerian motion.  Both the mass and motion of the centrifugally-supported disks produced in the simulations by Machida et al. (2011a; 2011b) do not match those inferred for the Keplerian circumstellar disks of Class~I protostars.  Nevertheless, the fact that Machida et al. (2011a; b) were able to successfully produce centrifugally-supported disks around protostars may provide important insights on how Keplerian circumstellar disks form around protostars.  Li, Krasnopolsky, \& Shang (2011) argue, however, that Machida et al. (2011a; b) do not properly handle the accumulation of magnetic flux at small radii because their simulations lack sufficient spatial resolution ($i.e.$, numerical diffusivity), artificially weakening the efficiency of magnetic braking.

In a recent paper \citep[][hereafter Paper~I]{tak12}, we reported the detection of a Keplerian circumbinary disk around the Class~I system L1551~NE.  Located in the Taurus molecular cloud at a distance of $\sim$140~pc \citep{eli78}, L1551~NE exhibits two 3.6-cm radio continuum sources (comprising free-free emission from protostellar jets) with a projected separation of $\sim$70 AU at a position angle of $\sim$300$\degr$ \citep{rei02}.
Reipurth et al. (2002) referred to the south-eastern source as ``Source A" and the north-western source as ``Source B," a nomenclature that we have retained.  Observations in the near-infrared revealed that Source A drives a pair of collimated [Fe II] jets along the north-east to south-west direction at a position angle of $\sim$60$\degr$ \citep{rei00,rei02,hay09}. Source B is located at the origin of an extended ($\sim$2000 AU) NIR reflection nebula corresponding to an outflow cavity \citep{rei00,rei02,hay09}.  Source B must therefore have drove a stronger outflow in the recent past that carved out the observed outflow cavity.  Our observation of L1551~NE reported in Paper~I was made with the SubMillimeter Array (SMA)\footnote{The SMA is a joint project between the Smithsonian Astrophysical Observatory and the Academia Sinica Institute of Astronomy and Astrophysics and is funded by the Smithsonian Institution and the Academia Sinica.} in its extended configuration and in the 0.9-mm continuum as well as $^{13}$CO~(3--2) and C$^{18}$O~(3--2) lines.  These observations revealed compact (unresolved) dust components associated with each binary component suggestive of circumstellar disks, as well as a circumbinary disk of radius $\sim$300~AU exhibiting Keplerian motion.  The mass (in both molecular gas and dust) of the circumbinary disk as estimated from the 0.9-mm dust continuum is $\sim$0.03 $M_{\odot}$.  From model fitting to the C$^{18}$O velocity channel maps, we inferred a disk inclination of $\sim$62$\degr$, position angle for its major axis of $\sim$167$\degr$, and total mass for the enclosed binary protostellar components of $\sim$0.8 $M_{\odot}$; i.e., the circumbinary disk is about an order of magnitude less massive than the protostellar system contained within.  Evidently, by or during the Class~I stage, Keplerian disks have formed or can form not only around individual protostars, but also around the binary components of a protostellar system.

Due both to its relative proximity and the relatively large size of its circumbinary disk, L1551~NE presents a good opportunity to study the kinematics of material just beyond the centrifugally-supported disk of a protostellar system.  In this way, we hope to better inform efforts to model how such disks form.  In this paper, we report follow-up observations of L1551~NE with the SMA in the sub-compact and compact configurations that cover spatial scales as large as $\sim$2000 AU, which in our previous observation (Paper~I) were strongly if not completely resolved out by the interferometer.  Our goals are to better define the outermost extent of the circumbinary disk exhibiting Keplerian motion (hereafter the Keplerian radius), and to infer the dynamics of material immediately beyond this portion of the disk.  In $\S$2, we describe our observations and data reduction.  In $\S$3, we present our results for the spatial and velocity distribution of the C$^{18}$O~(3--2) emission from the new data, as well as the combined new and old (from Paper~I) data.  In $\S$4, we model the observed gas kinematics to infer the centrifugally-supported extent of the circumbinary disk.  We also report the detection of infalling material with little if any angular momentum beyond the Keplerian radius, and discuss the implications of our findings for the formation of the Keplerian circumbinary disk in L1551~NE.  In $\S$5, we provide a concise summary of the results and our interpretation, and offer a few thoughts for future work.

\section{Observations and Data Reduction}

We observed L1551 NE in the C$^{18}$O $J$=3--2 (rest frequency of 329.3305453 GHz) line with the SMA
\citep{ho04}
in its compact configuration on 2011 Dec 28 and its subcompact configuration on 2012 Jan 14.  
%Details of the SMA are described by Ho et al. (2004).
In Table \ref{tbl-1}, we summarize the relevant parameters of the telescope and the calibrators used in our observations.  The frequency resolution employed in these observations, corresponding to a velocity resolution of $\sim$$0.185 {\rm \ km \ s^{-1}}$, is a factor of 2 higher than that used in our previous observations when SMA was in its extended configuration (Paper~I).  The raw visibility data were calibrated and flagged with MIR, which is an IDL-based data reduction package \citep{sco93}.
% The estimated uncertainty in the absolute flux calibration is $\sim$30$\%$.
We then used the software package Miriad \citep{sau95} to Fourier-transform and CLEAN the calibrated visibility data to produce the final images.  The images made from the newly-acquired data in the compact and subcompact configurations added together are henceforth referred to as the ``new'' images.  We also combined our newly-acquired data with our previous data taken in the extended configuration (Paper~I) to produce ``combined'' images.  
%Because the velocity resolution of the new data ($\sim$$0.185 {\rm \ km \ s^{-1}}$) is a factor of 2 higher than that of our previous data taken with the extended configuration.
The resultant image parameters are listed in Table \ref{tbl-2}.  The minimum projected baseline length is $\sim$$6.2 {\rm \ k\lambda}$,
so that 10 - 50$\%$ of the peak flux can be recovered for a Gaussian emission distribution
with a full-width half-maximum (FWHM) of 15$\arcsec$ - 27$\arcsec$
\citep{wil94}, and a larger fraction still for a Gaussian emission distribution with a smaller FWHM.
% and for a Gaussian emission distribution with a FWHM of $\sim$15$\arcsec$
% ($\sim$2100 AU), the peak flux recovered is $\sim$50$\%$ of the
% peak flux of the Gaussian \citep{wil94}.

\placetable{tbl-1}
\placetable{tbl-2}

\section{Results}

Figure~\ref{mom0} shows the combined C$^{18}$O (3--2) integrated intensity map of L1551 NE.  As in Paper~I, we find an elongated feature along the north-west to south-east direction centered approximately at the position of Source A.  The major and minor axes of this elongated feature match well with the corresponding axes of the circumbinary disk reported in Paper~I.  As explained in that paper, this circumbinary disk has its major axis at a position angle of 347$\degr$ and minor axis at 77$\degr$, as indicated by the dashed lines in Figure \ref{mom0}.  The major axis of the circumbinary disk is approximately perpendicular to the axis of the [Fe II] jets driven by Source A \citep{hay09}.  By comparison with the circumbinary disk reported in Paper~I, the elongated feature detected here has an outer dimension of $\sim$1000 AU $\times$ 800 AU that is significantly larger than the dimensions of the circumbinary disk of $\sim$600 AU $\times$ 300 AU (made from data taken only with the extended configuration).

%% Line profiles and missing fluxes.....--> Still significant missing fluxes....

Figure \ref{ch18} shows the combined velocity-channel maps in C$^{18}$O (3--2).  From the symmetry in velocity space evident in these maps and the Position-Velocity (P-V) diagrams described below, we hereafter adopt a systemic velocity for L1551~NE of $V_{LSR}$ = 6.9 km s$^{-1}$. Over the blueshifted range ($V_{LSR}$ = 4.2--6.1 km s$^{-1}$)
the C$^{18}$O emission is located predominantly to the north
(hereafter referred to as the ``high-velocity blueshifted'' component),
whereas over the redshifted range (7.9--9.4 km s$^{-1}$) the emission is located to the south (``high-velocity redshifted'' component), of the protostellar system.
Over the redshifted range of 7.2--7.5 km s$^{-1}$ (``low-velocity redshifted'' component), there is strong emission to the south as seen for the high-velocity redshifted component but also relatively weak and extended emission to the north.
% while there is no detectable blueshifted counterpart to the south.
Figure \ref{chlv} shows velocity-channel maps over a smaller velocity range of 6.6--7.2 km s$^{-1}$ but at a higher velocity resolution (0.185 km s$^{-1}$) made from just the new data.
% The northern extension at the low redshifted velocity is also seen in the new data.
In these higher velocity-resolution channel maps, 
%which also have a higher brightness-temperature sensitivity than the channel maps at a higher angular (but lower velocity) resolution***CORRECT*** shown in Figure \ref{ch18}, extended blueshifted emission also can be seen to the south of L1551~NE at 6.8 km s$^{-1}$ as well as the abovementioned northern redshifted extension.  
the redshifted emission close to the systemic velocity can be seen to be more strongly weighted towards the east of the protostellar system whereas the blueshifted emission more strongly weighted towards the west.  Thus, there appears an east (red) to west (blue) velocity gradient at low velocities close to the systemic velocity.

Figure \ref{br18} shows the combined ($left \ column$) and new maps ($right \ column$) integrated over the high-velocity blueshifted and redshifted components ($top \ row$), the low-velocity blueshifted component ($middle \ row$), and the low-velocity redshifted component ($bottom \ row$).  In the high-velocity components, the blueshifted and redshifted emission are well-separated and located to the north and south, respectively, of the protostellar system.  The orientation of these blueshifted and redshifted components and hence the orientation of their velocity gradient match well with the orientation of the major axis of the circumbinary disk, suggesting that the high-velocity components primarily trace the Keplerian circumbinary disk as found in Paper I.  In the low-velocity components, blueshifted emission is seen towards the north of the binary components as in the high-velocity components; on the other hand, redshifted emission is seen not just to the south but also to the north of the binary components. 
% {[\bf the referee is not objecting to our use of on the other hand, but just as a matter of writing style by using it too many times.  i have no problem using on the other hand as many times as necessary in a manuscript if it makes the arguments CLEAR.   i do not object to using an alternative phrase, EXCEPT when it makes things UNCLEAR.  this is such an example here.  keep the original phrase.  clarity beats style every time!]}
Along the minor axis, the intensity of the low-velocity redshifted emission is weighted towards the east of the protostellar system and that of the low-velocity blueshifted emission weighted towards the west, indicating an east (red) to west (blue) velocity gradient along the minor axis at velocities close to the systemic velocity as mentioned earlier.  This velocity gradient along the minor axis is clearly seen in the P-V diagram of C$^{18}$O (3--2) as we describe next.

Figure \ref{pv} ($left$ $column$) shows the P-V diagrams in C$^{18}$O, derived from the combined images, along the major ($upper$ $row$) and minor ($lower$ $row$) axes of the circumbinary disk and passing through the position of Source A.  Along the major axis, the high-velocity blueshifted and redshifted components are individually separated and located to the north and south, respectively, of Source A.  The emission at higher velocities is concentrated closer to Source A than that at lower velocities, a characteristic feature of Keplerian disks as illustrated by the blue curves corresponding to the Keplerian rotation-curve derived in Paper I for the circumbinary disk.  In addition to the features tracing Keplerian rotation, around the systemic velocity there also is a spatially-extended redshifted feature toward the north of the Source A (not seen previously in Paper~I); this feature arises from the low-velocity redshifted component towards the north of the binary components mentioned above.  Along the minor axis, the P-V diagram exhibits a clear velocity gradient at velocities relatively close to the systemic velocity (again not previously seen in Paper~I) as indicated by the arrows in the lower left panel of Figure \ref{pv}.  By contrast, no velocity gradient is evident along the minor axis at higher velocities.  The velocity gradient seen at relatively low velocities along the minor axis therefore reflects the motion of the eastern low-velocity redshifted and western low-velocity blueshifted components.  For purely Keplerian rotation, no velocity gradient should be evident along the minor axis, and indeed no velocity gradient is seen along the minor axis for the high-velocity components.  The observed velocity gradient along the minor axis for emission at relatively low velocities, seen here for the first time from observations sensitive to larger-scale structures than in Paper~I, therefore indicates motion not related to Keplerian motion.  In the next section, we will discuss the nature of this newly-discovered component.

%%Our new SMA observations of L1551 NE in the C$^{18}$O emission with the subcompact and compact configurations have therefore revealed a new component in addition to the Keplerian disk component found with the SMA extended configuration in Paper I.

\section{Discussion}
\subsection{Nature of the Low-Velocity Components}
% \section{Analysis}

As described in the previous section, we find that the C$^{18}$O (3--2) emission surrounding L1551~NE exhibits two distinct velocity components: (i) high-velocity blueshifted and redshifted components that are more spatially-extended along the north-south than east-west directions, and which show a velocity gradient along the more spatially-extended north-south direction.  These components trace a Keplerian circumbinary disk as identified in Paper~I; and (ii) low-velocity blueshifted and redshifted components that also are more spatially-extended along the north-south than east-west directions, but which unlike the high-velocity blueshifted and redshifted components show a velocity gradient along the less spatially-extended east-west direction.
%{\bf [this is the key phrase; more spatially-extended along the north-south than east-west]}
Below, we consider two alternative explanations and offer our opinion for the likely nature of the low-velocity components.

% The low-velocity components are unlikely to correspond to a molecular outflow.
One possible interpretation for the low-velocity components is a molecular outflow.  From observations with the JCMT, Moriarty-Schieven et al.~(1995) found that the CO~(3-2) emission around L1551~NE exhibits blueshifted and redshifted line wings (at velocities $\gtrsim$2.5 km s$^{-1}$ from the systemic velocity) in spectra taken towards the west and east, respectively, of the protostellar system. They attributed the emission in the line wings to a molecular outflow driven by L1551 NE. Furthermore, near-IR [Fe II] observations of L1551 NE show knotty jets driven by Source A that are blueshifted to the south-west and and redshifted to the north-east \citep{rei00,rei02,hay09}.  Thus, the line wings in CO~(3-2) as well as jets driven by Source~A share the same sense in velocity gradient in the sky as the low-velocity C$^{18}$O (3--2) emission reported above.  If the low-velocity blueshifted and redshifted C$^{18}$O (3--2) components indeed trace
a molecular outflow, then based on their projected velocity ($\sim$0.5 km s$^{-1}$; see black arrows in Figure \ref{pv}) and projected separation
from Source A ($\sim$2$\arcsec$, corresponding to $\sim$280~AU),
and assuming that the outflow emerges perpendicular to the Keplerian circumbinary disk (inclination of $i$ $\sim$ 62$\degr$; see Paper~I), the deprojected outflow velocity is then $v_{flow}$ $\sim$ 0.5 km s$^{-1}$ / $\cos i$ $\sim$ 1.0 km s$^{-1}$ and the deprojected spatial separation of the individual outflow components $l_{flow}$ $\sim$ 280~AU / $\sin i$ $\sim$ 320 AU from Source A. By comparison, at their separation from the protostellar system, the escape velocity is
\begin{equation}
v_g = \sqrt{\frac{2GM_{\star}}{l_{flow}}},
\end{equation}
where $G$ is the gravitational constant and $M_{\star}$ is the mass of the protostellar system (= 0.8 $M_{\odot}$; Paper I).  We find $v_g$ $\sim$ 2.1 km s$^{-1}$, a factor 2 higher than the deprojected outflow velocity mentioned above of $v_{flow}$ $\sim$ 1.0 km s$^{-1}$.
The low-velocity components are therefore bound to the protostellar system.  Indeed, the low-velocity components reach much
lower blue- and redshifted velocities ($\lesssim  \pm 0.5$ km s$^{-1}$ from the systemic velocity) than the CO (3--2) outflow
observed with the JCMT, even though the JCMT observations probe regions much further away from L1551~NE than that
spanned by the low-velocity components. Furthermore, contrary to expectations for a molecular outflow driven along the east-west directions, the low-velocity components are even more spatially-extended along the north-south than east-west directions.
%{\bf Therefore, the low-velocity C$^{18}$O emission is unlikely to trace the outflowing gas being ejected from the gravitational potential of the central protostars.  Furthermore, the low-velocity components are elongated along the north-south direction (as shown earlier in Figure \ref{chlv}) and thus perpendicular to the direction of their velocity gradient and the associated outflow. Molecular outflows are often observed as bipolar, elongated features along the direction of the velocity gradients, and the observed elongation of the low-velocity C$^{18}$O emission is inconsistent with that of outflows.  These results show that the low-velocity blueshifted and redshifted C$^{18}$O (3--2) components are unlikely to be originated from the outflow.}

   % Furthermore,
% as seen in Figures \ref{chlv} and \ref{br18}, the low-velocity components appear elongated perpendicularly
% to the outflow axis.
% and the extent ($\sim$1000 AU) is larger than that of the circumbinary disk ($\sim$600 AU).

Alternatively, we consider the possibility that the low-velocity components correspond to gas co-planar with but lying just beyond the Keplerian radius of the circumbinary disk.  As described in $\S$3, the major axis of the circumbinary disk is orthogonal to the projected axes of the [Fe II] jets from Source A; if indeed the plane of the circumbinary disk is perpendicular to the axis of the jets, the western part of the disk plane would then correspond to the far side and the eastern part of the disk plane to the near side.  Correspondingly, the western low-velocity blueshifted component would then be located on the far side and the eastern low-velocity redshifted component the near side.  The low-velocity components would therefore be infalling towards the circumbinary disk along the disk plane.  
%Such a situation also has been seen in C$^{18}$O (2--1) around the single class~0 protostar B335, which is surrounded by disk-like structure with no detectable rotation but which also \citep{yen10}.
The infall velocity, $v_{inf}$, can be roughly estimated from the velocity gradient seen for the low-velocity components along the minor axis of the circumbinary disk in the P-V diagram of Figure \ref{pv}.  At the opposing locations of the low-velocity components, the velocity shift across the minor axis is $\sim$1.0 km s$^{-1}$ (see arrows in Figure \ref{pv}).  At an inclination of $i$ $\sim$ 62$\degr$ for the circumbinary disk, the infall velocity along the disk plane is therefore $v_{inf}$ $\sim$ (1.0 km s$^{-1}$/2) / sin 62$\degr$ $\sim$ 0.6 km s$^{-1}$.

To see if the above interpretation offers a good fit to the observed P-V diagrams, we have generated P-V diagrams based on a simple toy model to compare with the observed P-V diagrams.  Our model comprises a geometrically-thin disk as was adopted to model the Keplerian circumbinary disk described in Paper~I, but with a velocity field now described by:
% The centroid line-of-sight velocity at each position of the geometrically-thin disk
% in right ascension ($\equiv \alpha$) and declination ($\equiv \delta$) with respect to the
% disk center ($\equiv$ $v_{LOS} (\alpha,\delta)$)
% can be expressed as;
% \begin{equation}
% v_{LOS} (\alpha,\delta) = v_{sys} + v_{rot} (r) \cos(\Phi-\theta) + v_{rad} (r) \sin(\Phi-\theta),
% \end{equation}
% where
% \begin{equation}
% \Phi = \arctan(\frac{\alpha}{\delta}),
% \end{equation}
% \begin{equation}
% r = \sqrt{(\frac{x}{\cos i})^2 + y^2},
% \end{equation}
% \begin{equation}
% x = \alpha \cos(\theta) - \delta \sin(\theta),
% \end{equation}
% \begin{equation}
% y = \alpha \sin(\theta) + \delta \cos(\theta).
% \end{equation}
% In the above expressions $v_{sys}$ is the systemic velocity, $r$ is the radius,
% $\theta$ is the position angle of the disk major axis, $i$ is the disk inclination angle
% from the plane of the sky,
% and $x$ and $y$ are coordinates of the disk along the minor and major axes, respectively.
% $v_{rot} (r)$ and $v_{rad} (r)$ denote the rotational and infalling velocities of the disk
% as a function of the radius.
% Here we incorporate two distinct velocity structures in the disk; one is a Keplerian disk
% with the outer radius $r_{kep}$, and the other outer infalling ``ring'' without any rotation
% surrounding the Keplerian disk;
%
\begin{mathletters}
\begin{eqnarray}
v_{rot}(r)=\sin{i}\sqrt{\frac{GM_{\star}}{r}} \hspace{0.5cm} {\rm and} \hspace{0.5cm} v_{rad} (r) = 0 \hspace{0.5cm} {\rm for} \hspace{0.5cm} r \leq r_{kep}, \\
v_{rot}(r)=0 \hspace{0.5cm} {\rm and} \hspace{0.5cm}  v_{rad} (r) = v_{inf} \hspace{0.5cm} {\rm for} \hspace{0.5cm} r > r_{kep},
\end{eqnarray}
\end{mathletters}
\noindent where $r$ is the radius along the disk, $i$ is the disk inclination, $v_{rot} (r)$ the rotational velocity of the disk (i.e., the portion corresponding to the Keplerian circumbinary disk), $v_{rad} (r)$ the radial (infall) velocity of the disk (i.e., the portion corresponding to the low-velocity components), $r_{kep}$ the Keplerian radius, and $v_{inf}$ an adopted constant infall velocity beyond the Keplerian radius.  
% $G$ is the gravitational constant, $M_{\star}$ is the mass of the central star,
Like in Paper~I, we adopt $M_{\star}$ = 0.8 $M_{\odot}$, $i$ = 62$\degr$, a disk position angle of $\theta$ = 167$\degr$ with  the dynamical center at Source A, and an internal velocity dispersion $\sigma_{gas}$ = 0.4 km s$^{-1}$.  To model the intensity distribution of the disk, we adopt an intensity profile that resembles the intensity distribution of the C$^{18}$O emission as derived from the moment 0 map (The C$^{18}$O emission shows no evidence for candidate circumstellar disks associated with the two binary components as seen in the continuum at 0.9~mm and reported in Paper~I; we therefore believe that the observed C$^{18}$O is dominated by emission from the circumbinary disk.). That is, we fitted two-dimensional Gaussian structures to the observed moment 0 map of Figure~\ref{mom0} to represent the intensity distribution of the disk; we found that a minimum of three Gaussian components were required to produce a satisfactory fit, comprising two relatively compact Gaussians each centered on the two peaks seen in the moment 0 map, along with a much broader Gaussian to capture more extended emission surrounding the two peaks.  By varying the Keplerian radius ($r_{kep}$) and the infall velocity beyond the Keplerian radius ($v_{inf}$), we found from trial and error a model P-V diagram that we judged (by eye) to provide a satisfactory fit to the observed P-V diagrams.  To generate the model P-V diagrams, first each trial velocity field for the disk was converted into the line-of-sight velocity on the disk plane, and then model velocity-channel maps created from the line-of-sight velocity map, the model moment 0 map, and the assumed internal velocity dispersion ($\sigma_{gas}$).  The model P-V diagrams were then generated from the model velocity-channel maps.

The model described above ignores a potential envelope component (corresponding to the parental dense core of L1551~NE) in which the the disk component is embedded, and which if not completely recovered in our SMA observations introduces artifacts in the maps constructed from the available data and also the velocity field derived from these maps (e.g., P-V diagrams).  To simulate the effect such an envelope would have on our model P-V diagrams, we have added an envelope component to the disk component described above so that our final model comprises  a disk embedded in a spherically-symmetric envelope.  Using the $uv$-coverage attained in our SMA observations, we made mock observations of the model velocity-channel maps
having different $r_{kep}$ and $v_{inf}$ for the disk component
as described above and the parameters for the envelope component as described below to generate model visibilities, and then Fourier-transformed and CLEANED the model visibilities to produce simulated-observed model velocity-channel maps.  The latter maps are then used to derive simulated-observed model P-V diagrams.  We decided upon the parameters of the model envelope component in the following manner.  A C$^{18}$O (3--2) spectrum taken with the CSO towards the center of L1551~NE exhibits two distinct components: one centered at $V_{\rm LSR}$$\sim$6.7 km s$^{-1}$ that has a narrower velocity width of $\sim$0.68 km s$^{-1}$, and the other centered at $V_{\rm LSR}$$\sim$7.0 km s$^{-1}$ that has a broader velocity width of $\sim$2.2 km s$^{-1}$ \citep{ful02}.  The central velocity and velocity range spanned by the broader component are similar to those of the C$^{18}$O (3--2) emission that we mapped with the SMA.  Indeed, a comparison of the SMA C$^{18}$O (3--2) spectrum with the CSO C$^{18}$O (3--2) spectrum shows that virtually all of the flux in the CSO spectrum is recovered in our SMA observation in the ($V_{\rm LSR}$) velocity range $\sim$3.9--5.4 km s$^{-1}$ and $\sim$8.7--9.4 km s$^{-1}$ (although the poor S/N of the CSO spectrum in the line wings prevents us from precisely estimating the amount of the recovered flux). By contrast, around the systemic velocity, we miss in our SMA observations a significant amount of flux present in the CSO spectrum.  In the lower blueshifted-velocity range 6.4--6.8 km s$^{-1}$, the amount of the recovered flux is only 7--8$\%$ of the CSO flux, whereas in the lower redshifted-velocity range 7.2--7.5 km s$^{-1}$, the amount of recovered flux is 30--40$\%$.  Much the missing flux corresponds to the narrower component in the CSO spectrum described above; this component must therefore correspond to a spatially-extended envelope.  Unfortunately, there are no available single-dish maps of the C$^{18}$O (3--2) emission around L1551 NE and so the structure and the kinematics of this envelope are unclear.  Observations of L1551 NE in the 850 $\micron$ dust continuum emission with SCUBA on the JCMT show that the size of the surrounding dusty envelope is $\sim$9$\farcs$8$\times$8$\farcs$8 \citep{mor06}.  We therefore anticipate that the extent of the envelope in C$^{18}$O (3--2) is unlikely to be much larger than the CSO beam of $\sim$23$\arcsec$.  For the envelope component in our model, we have therefore adopted a spherically-symmetric Gaussian profile with a cross-sectional size at FWHM that is the same as that of the CSO beam, together with a central velocity ($V_{\rm LSR}$ = 6.7 km s$^{-1}$), line width (0.68 km s$^{-1}$), and total intensity (55.5~Jy) having the same values as the corresponding parameters for the narrower component in the CSO C$^{18}$O spectrum.

%Thus, we created a 2-dimensional Gaussian image with the FWHM of 23$\arcsec$ and the total flux equal to that of the narrow component in the CSO C$^{18}$O spectrum (= 55.5 Jy). Then the line width of 0.68 km s$^{-1}$ and the central velocity of $V_{\rm LSR}$ = 6.7 km s$^{-1}$ derived from the CSO spectrum are applied to the Gaussian image, and the velocity channel maps of the model envelope were created. The image cubes of the disk and the envelope components were co-added, and sampled with the $uv$ sampling of the real SMA observations to produce the model visibilities. Then the model visibilities were Fourier-transformed and CLEANed in the same way as the real SMA visibility data, to make the model P-V diagrams.

Figure~\ref{pv} (middle column) shows the simulated-observed model P-V diagrams along the major (upper row) and minor (lower row) axes with $r_{kep}$ = $\infty$ so that the model disk exhibits Keplerian motion only (labeled the ``Keplerian only model"), and Figure \ref{pv} (right column) the corresponding simulated-observed model P-V diagrams but now with $r_{kep}$ = 300 AU and $v_{inf}$ = 0.6 km s$^{-1}$ so that the model disk exhibits infall (with no rotation) beyond the Keplerian radius (labeled the ``infall $\rightarrow$ Keplerian model").  (Recall that the intensity profile of the observed moment 0 map constrains the outer radius of the model disk that contributes significantly to the emission.)  We also made simulated-observed model velocity-channel maps without the envelope component to check whether the inclusion of this component changes the simulated-observed model P-V diagrams compared with those shown in Figure~\ref{pv}, but did not find any important differences.  For comparison, the best-fit Keplerian-rotation curve derived in Paper I is drawn in all the P-V diagrams along the major axis (blue curves in Figure~\ref{pv}), and the curve for free-fall onto a central system mass of $M_{\star}$ = 0.8 $M_{\odot}$ drawn in all the P-V diagrams along the minor axis (red curves). In the P-V diagram along the major axis, the model that invokes only Keplerian rotation in the disk (Keplerian only model) shows that the blueshifted and redshifted components lie at distinct opposite quadrants in the P-V plane, such that  emission at higher velocities is located closer to the center than that at lower velocities; these characteristics are similar to those seen in the P-V diagrams for the observed high-velocity blueshifted and redshifted components.  This model, however, cannot reproduce the observed spatially-extended low-velocity component to the north of Source A in the P-V diagram along the major axis.  Furthermore, in the P-V diagram along the minor axis, the model incorporating only a Keplerian disk does not reproduce the observed velocity gradient at velocities relatively close to the systemic velocity. If we truncate the Keplerian disk at $r_{kep}$ $\sim$300 AU and add an outer infalling component along the disk plane with $v_{inf}$ $\sim$0.6 km s$^{-1}$ (infall $\rightarrow$ Keplerian model), we can reproduce the extended low-velocity component in the P-V diagram along the major axis, as well as the velocity gradient in the P-V diagram along the minor axis at velocities relatively close to the systemic velocity.  
%Note that the shorter arm of the cross-shaped feature would be vertical in the P-V diagram along the minor axis if the portion of the disk beyond that exhibiting Keplerian rotation exhibited no infall (i.e., $v_{inf}$ = 0), just as in Figure \ref{pv} (middle column, lower panel) for a disk exhibiting Keplerian rotation (i.e., no infalling motion) only {\bf (Keplerian P-V is cross shaped or not ??)}.
%In the P-V diagram of the infall $\rightarrow$ Keplerian model along the minor axis, the vertical stroke of the cross shape becomes tilted, thus resembles the observed velocity gradient in the low-velocity region. Here, the outer radius of the infalling component is constrained by the observed moment 0 map (Figure \ref{mom0}), since we assumed that the moment 0 map of the model is the result of the three-components Gaussian fitting to the observed moment 0 map.

In summary, our simple toy model of a circumbinary disk exhibiting a combination of Keplerian motion in its inner region and infall beyond its Keplerian radius provides an excellent fit to the gross spatial-kinematic morphology of the observed C$^{18}$O
emission as captured in the P-V diagrams (irrespective of whether of not the disk is embedded in a spatially-extended envelope).  We therefore believe that the low-velocity components newly detected here in observations sensitive to larger-scale structures than those in Paper~I most likely trace infalling gas located just beyond the Keplerian circumbinary disk first detected in Paper~I.
%We suggest that our follow-up SMA observations with shorter $uv$-spacings than that of Paper I have revealed the outer infalling gas toward the inner Keplerian circumbinary disk found in Paper I. In other words, we have found where the outer infalling region ``transforms'' into the inner Keplerian circumbinary disk.
We note that the model (constant) infall velocity beyond the Keplerian radius of $v_{inf}$ $\sim$0.6 km s$^{-1}$ (arrows in Figure~\ref{pv}) is smaller than the expected free-fall velocity (red curves in Figure~\ref{pv}) onto a central stellar mass (for both binary components combined) of $M_{\star}$ = 0.8 $M_{\odot}$ over the radial range from the protostellar system (Source A) plotted in Figure \ref{pv}; e.g., at a radius of $\sim$300~AU (the Keplerian radius), the predicted free-fall velocity is $\sim$2.2 km s$^{-1}$.  Instead, the model infall velocity of $\sim$0.6 km s$^{-1}$ at $r$=300 AU implies a central stellar mass of only $\sim$0.06 $M_{\odot}$, which is an order of magnitude smaller than the formal value for the mass of the protostellar system estimated from the Keplerian motion of the circumbinary disk of 0.8$^{+0.6}_{-0.4}$ $M_{\odot}$ (uncertainty corresponding to the FWHM of the $\chi$-squared curve; see Paper I).  The smaller-than-expected infall velocity may indicate that the infalling material is decelerated (presumably by gas pressure, if not also turbulence) as it reaches close to the Keplerian disk. On the other hand, we emphasize that, in our toy model, we did not include any rotation in the infalling region.  We have attempted a number of models that connect the rotational profile of the Keplerian disk at $r_{kep}$ to the outer infalling region using a $v_{rot}(r) \propto r^{-1}$ rotational profile, but such velocity profiles cannot reproduce the low-velocity redshifted component to the north apparent in the observed P-V diagram along the major axis.  This failure to connect the Keplerian to the infalling region using a sensible rotational profile may indicate an abrupt increase in the rotational velocity progressing inwards from the infalling region to $r_{kep}$.
%In the next subsection we will discuss these results and the possible interpretations.

% Such a signature of the rotational profile can be seen in
% theoretical simulations of disk formation \citep{mac10,ma11b}, although those theoretical simulations
% cannot fully reproduce observed Keplerian disks (see Paper I and discussion below).
%
% Further high-spatial dynamic range, high-sensitivity
% observations of the ``transitional'' region
% from the outer infalling envelope to the inner Keplerian disk with ALMA
% should unveil the physics of the transition and constrain theories of disk formation.

\subsection{On the Formation of Keplerian Disks}

Although theoretical models have so far only investigated the possible formation of Keplerian circumstellar disks around single protostars, we anticipate the same models to reach their same conclusions on the formation of Keplerian circumbinary disks around the closely-separated components of a binary protostellar system like L1551~NE.  Indeed, the mass ($\sim$0.03--0.12 $M_{\odot}$; Paper~I) and size (outer radius $\sim$300~AU) of the Keplerian circumbinary disk around L1551~NE, a Class~I object, is comparable to the masses and sizes of circumstellar disks observed around single Class~I protostars.

As mentioned in Section~1, despite the ever increasing evidence for Keplerian circumstellar disks around Class I protostars, many theoretical models continue to have difficulty forming such disks around protostars.  
%Although there are an increasing number of observational findings of large-scale ($r$ $\gtrsim$ a few $\times$ 100 AU) Keplerian disks around Class I protostars \citep{bri07,lom08,jor09,tob12,yen13}, including the present study, it has been still difficult to theoretically form such Keplerian disks from protostellar envelopes.
Mellon \& Li (2009) and Li, Krasnopolsky, \& Shang (2011) have conducted thorough 2-D axisymmetric simulations of the collapse of rotating and magnetized singular isothermal cores, in which they included non-ideal MHD effects such as ambipolar diffusion, the Hall effect, and Ohmic dissipation.  Their detailed investigations show that even in the case of moderately magnetized cases (mass-to-flux ratio $\lambda$ $\lesssim$10), magnetic braking is strong enough to remove essentially all of the angular momentum of the material that accretes onto the central object and thus suppress the formation of any sizable ($\gtrsim$ several AU) centrifugally-supported disks.  They argued that effects not included in their models such as outflow stripping of protostellar envelopes that anchor the magnetic field and/or enhanced magnetic diffusivity due to the turbulent-induced reconnection may be required to form Keplerian disks with radii $\gtrsim$ 100~AU. Machida et al. (2011a; b) have performed 3-D magnetohydrodynamic simulations of the evolution of Bonnor-Ebert spheres with different initial rotations and magnetic fields.  Under a broad range of conditions, they were able to successfully form centrifugally-supported disks with radii $\gtrsim$100 AU, although not until near or at the end of the main accretion phase.  Because the mass of the circumstellar disk in their model is considerably larger (by a factor of $\sim$2--100) than that of the central protostar (see also Vorobyov 2009), however, the circumstellar disk does not exhibit Keplerian motion and is subject to fragmentation.  As mentioned above, Li, Krasnopolsky, \& Shang (2011) argued that the simulations performed by Machida et al. (2011a; b) are severely affected by numerical diffusivity of the magnetic fields.

In the case of L1551~NE, our toy model that adequately reproduces the observed P-V diagrams comprises a circumbinary disk having motions dominated by Keplerian rotation inside a radius of $\sim$300~AU (the Keplerian radius) and infall (with an infall velocity considerably smaller than the free-fall velocity) outside this radius.  Such a break between predominantly rotational and predominantly infalling motions can be seen in the theoretical simulations by Machida et al. (2011a; 2011b), although as mentioned above these simulations produce much more massive centrifugally-supported circumstellar disks (which moreover do not exhibit Keplerian motion) than are actually observed.  The abrupt decrease in rotational velocity between the inner and outer regions of this disk indicates that angular momentum is effectively transported outwards, presumably by magnetic braking as invoked in theoretical models, beyond the Keplerian radius.  Within the Keplerian radius, we cannot easily quantify any deviation in the measured velocity field from Keplerian motion, although some loss of angular momentum is required for material to move through the circumbinary disk and eventually accrete onto the protostellar system.
%Our follow-up SMA observations of L1551 NE have found evidence for the outer infalling gas toward the inner Keplerian circumbinary disk. Furthermore, there is no obvious rotation in the infalling component, suggesting that the angular momentum at the outermost part of the Keplerian disk and that of the infalling component are not conserved. The magnetic braking is the most probable source of the removal of the angular momenta outside of Keplerian disk. At the same time, the presence of the well-developed Keplerian disk, where the gravity from the protostellar binary appears to dominate the local force, implies that magnetic field is negligible inside the Keplerian disk.

To explain the difference in the effectiveness at which angular momentum is transported outwards within and beyond the Keplerian radius in the circumbinary disk, we borrow ideas from Machida et al. (2011a; 2011b) as well as add our own.
One reason may be the difference in the plasma $\beta$ between the inner and outer regions of the disk, which is defined as,
\begin{equation}
\beta=\frac{8\pi P}{B^2},
\end{equation}
where $B$ and $P$ are the magnetic flux density and the thermal pressure, respectively \citep{ma11a,ma11b}.
Strictly speaking, plasma $\beta$ applies only to highly-ionized gas.  In the present situation of weakly-ionized and primarily molecular gas, the magnetic field is well coupled to the molecular gas through free electrons and ions, at least in directions perpendicular to the magnetic field.  Under such circumstances, the gas pressure refers not only to the ionized component, but also the neutral component; i.e., the total gas pressure, which in the present situation is dominated by the pressure of the molecular gas. To that we add the likelihood that the ionization fraction decreases dramatically between the outer and inner regions of the disk.  At smaller radii, the thermal pressure is likely to be higher because of the higher gas density and temperature; consistent with this idea, the inferred infall velocity is smaller than the expected free-fall velocity near the Keplerian radius, presumably because the infalling material is decelerated as it moves into regions of higher gas pressure.  In addition, ohmic dissipation of magnetic fields becomes increasingly more effective in denser regions closer to the protostar where cosmic rays do not penetrate as easily to ionize the molecular gas (Machida et al. 2011a; 2011b), and to which we add also because free electrons become increasingly depleted by adsorption onto dust grains in denser gas.  As a consequence, the plasma $\beta$ can increase and the ionization fraction decrease significantly towards smaller radii, and hence the control exerted by magnetic fields on the molecular gas become dramatically weaker at smaller radii. In this scenario, magnetic braking becomes ineffective within a certain (the Keplerian) radius, outside of which magnetic braking can operate effectively;
this scenario nonetheless
requires the initial infalling matter that forms the Keplerian disk to have retained
appreciable angular momentum.
The transition between good and poor magnetic control of the molecular gas may occur quite abruptly not only at a certain radius, but also at a particular stage in the evolution of a protostellar system as the mass and hence density of the circumbinary disk grows.
%Such a break between predominantly rotational and predominantly infalling motions has been found in the theoretical simulations by Machida et al. (2011a; 2011b), although, as mentioned above, these simulations produce much more massive Keplerian circumstellar disks than are actually observed.  
In our toy model, the transition between predominantly infalling to predominantly Keplerian motion in the circumbinary disk is infinitely sharp, a situation which is of course not physical; rather, our toy model suggests that this transition must occur over an annular region that is too narrow to be probed at the angular resolution of our observations.

%Finally, because it takes time for the mass and hence density of the circumbinary disk to grow, but for the drop in ionization fraction occur quite dramatically when the density in the inner region of the circumbinary disk 

%Further refinement of the density and temperature profiles as a function of the radius over the magnetic field distribution in the disk-forming region is likely required to theoretically reproduce the observed Keplerian disks.

Finally, we consider the possible future evolution of the Keplerian circumbinary disk in L1551~NE as it continues to accumulate mass from infalling material.
%L1551 NE is considered to be in the early Class I stage with the low bolometric temperature ($\sim$91 K; Froebrich 2005) and the surrounding dusty envelope ($\sim$0.39 $M_{\odot}$; Moriarty-Schieven et al. 2006).  Recently Tobin et al. (2012) have found a $r$ $\sim$150 AU Keplerian disk around L1527 IRS, a transitional object from the Class 0 to I stage, surrounded by a $\sim$1.0 $M_{\odot}$ envelope. These results suggest that large-scale ($r$ $\gtrsim$100 AU) Keplerian disks should be formed in the embedded protostellar phase. The infalling motion in the envelope surrounding the Keplerian disk should affect the future evolution of the Keplerian disk.
On the assumption of local thermodynamic equilibrium (LTE), an excitation temperature of C$^{18}$O (3--2) of 42~K \citep{mor94}, and a C$^{18}$O abundance of 1.7 $\times$ 10$^{-7}$ \citep{cra04}, we estimate a mass for the infalling component
detected in our observations of  $\sim$0.0023 $M_{\odot}$.
% This value is likely smaller than the mass of the Keplerian disk ($\sim$0.026 $M_{\odot}$; Paper I),
% even if the effect of the missing flux in the infalling component is taken into account.
For an infall velocity derived in our model of 0.6 km s$^{-1}$ at a radius of 300 AU,  the mass-infall rate is therefore \citep[e.g., see][]{tak07} $\sim$9.6 $\times$ 10$^{-7}$ $M_{\odot}$ yr$^{-1}$.  This value is, incidentally, comparable to the estimated mass-infall rate onto the Keplerian circumstellar disk of L1527 IRS of $\sim$6.6 $\times$ 10$^{-7}$ $M_{\odot}$ yr$^{-1}$ \citep{tob12}.  The envelope around L1551~NE has a mass (in both molecular gas and dust) of $\sim$0.39 $M_{\odot}$ as inferred from its dust continuum emission at 0.85~mm observed with the JCMT \citep{mor06}.  If this entire envelope drains onto the Keplerian circumbinary disk at the abovementioned mass-infall rate, the latter can continue to accumulate mass for the next $\sim$0.39 $M_{\odot}$ / 9.6 $\times$ 10$^{-7}$ $M_{\odot}$ yr$^{-1}$ $\sim$ 4.1 $\times$ 10$^{5}$ yr.  This timescale is comparable to the typical timescale estimated for the Class I evolutionary stage ($\sim$2.5 -- 6.7 $\times$ 10$^{5}$ yr; Hatchell et al. 2007).  Thus, the Keplerian circumbinary disk of L1551~NE may continue to grow in mass until the end of the main accretion phase.  Because the total mass of the Keplerian circumbinary disk would then become comparable to the present mass of the protostellar system, however, this disk may become unstable to fragmentation if all the mass that falls onto the Keplerian circumbinary disk remains there.  Of course, rather than growing too much in mass and fragmenting, much of the material in the Keplerian circumbinary disk may accrete onto the protostellar components and/or be ejected or entrained in an outflow. 
%Therefore, the Keplerian disk may become unstable during its growth from the surrounding material. Further accretion from the Keplerian disk to the central protostars is likely ongoing, unless most of the envelope material is dissipated by some mechanism such as the interaction with the outflows.

\section{Summary}

\begin{figure}
\epsscale{1.0}
\plotone{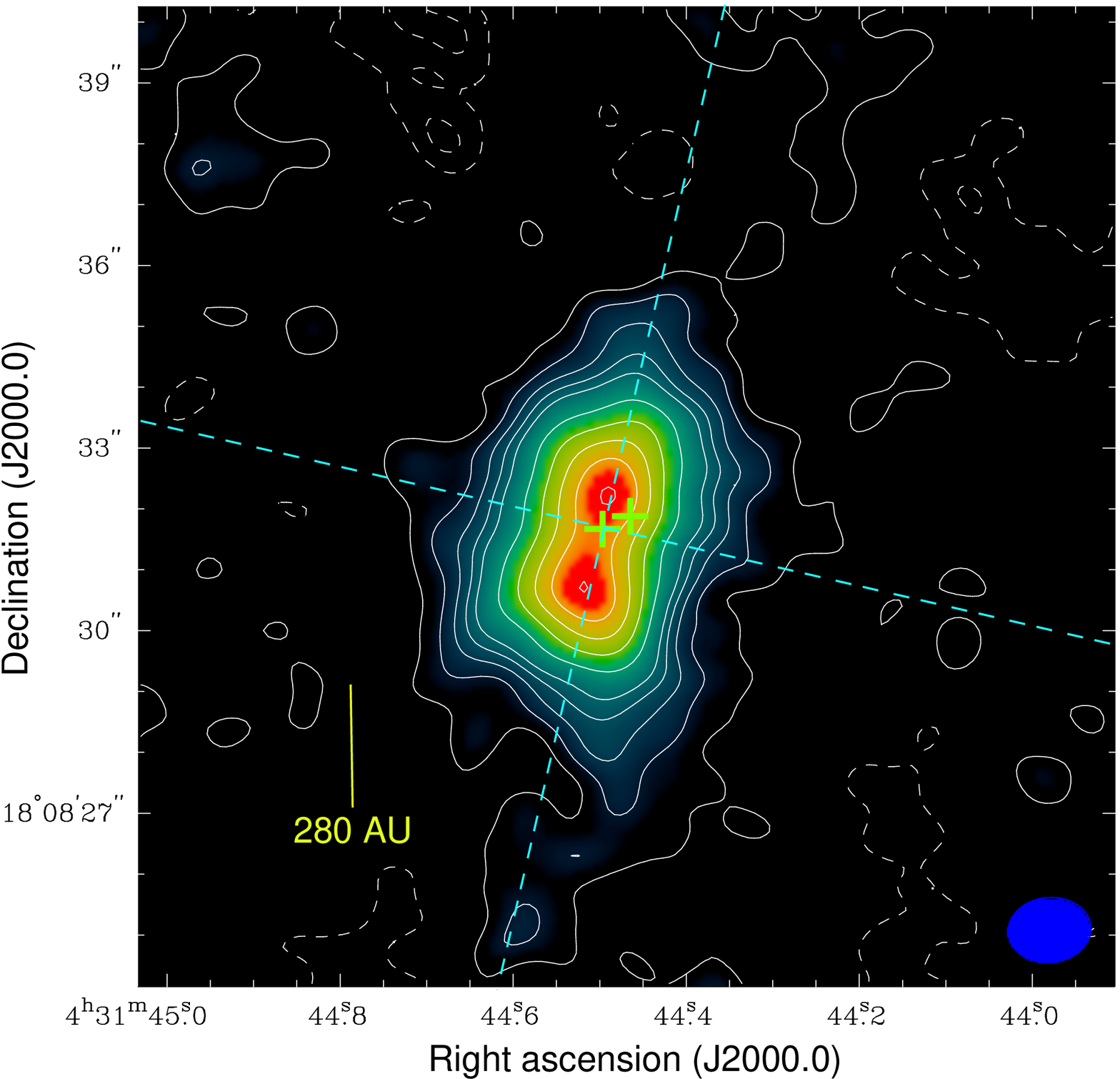}
\caption{Integrated C$^{18}$O (3--2) intensity (moment 0) map made by combining the newly-acquired data in the sub-compact and compact configurations with the previously published data (Paper~I) in the extended configuration of the SMA.  Contour levels are from 2$\sigma$ in steps of 2$\sigma$ until 10$\sigma$, and then in steps of 5$\sigma$ (1$\sigma$ = 0.119 Jy beam$^{-1}$ km s$^{-1}$).  The velocities used to make this map spans $V_{LSR}$ = 3.85--10.14 km s$^{-1}$.  Crosses indicate the positions of the binary components of L1551~NE (from Reipurth et al. 2002), with Source A located to the south-east and Source B located to the north-west.  Tilted dashed lines denote the major (P.A. = 347$\degr$) and minor (P.A. = 77$\degr$) axes of the Keplerian circumbinary disk discovered in Paper~I, and which passes through Source A.  The filled ellipse at the bottom-right corner corresponds to the synthesized beam as plotted at full-width half-maximum (FWHM) ($1\farcs36 \times 1\farcs06$ at P.A. = $-86\degr$.)
\label{mom0}}
\end{figure}

We have conducted follow-up observations of the binary protostellar system L1551 NE in C$^{18}$O (3--2) with the SMA in its subcompact and compact configurations.  We presented results made from the new data alone, as well those from combining the new data with that taken previously with the SMA in its extended configuration (Paper~I). Together, the combined data cover spatial scales spanning the range $\sim$140--2000 AU.  From the resulting maps and P-V diagrams, we found:

1.  An elongated feature around the binary protostellar system with an outer dimension of $\sim$1000 AU $\times$ 800 AU as shown in Figure \ref{mom0}.  The dimensions of this feature are significantly larger than those of the Keplerian circumbinary disk reported in Paper I of $\sim$600 AU $\times$ 300 AU imaged only with the extended configuration of the SMA.  The major axis of this feature matches well with the major axis of the Keplerian circumbinary disk, and is approximately perpendicular to the axis of the [Fe II] jets driven by Source A as well as the major axis of an outflow cavity centered on Source B.  This feature therefore traces a circumbinary disk or disk-like structure whose major axis, and presumably also equatorial plane, is perpendicular to outflows from the binary protostellar components.

\begin{figure}
\epsscale{1.0}
\plotone{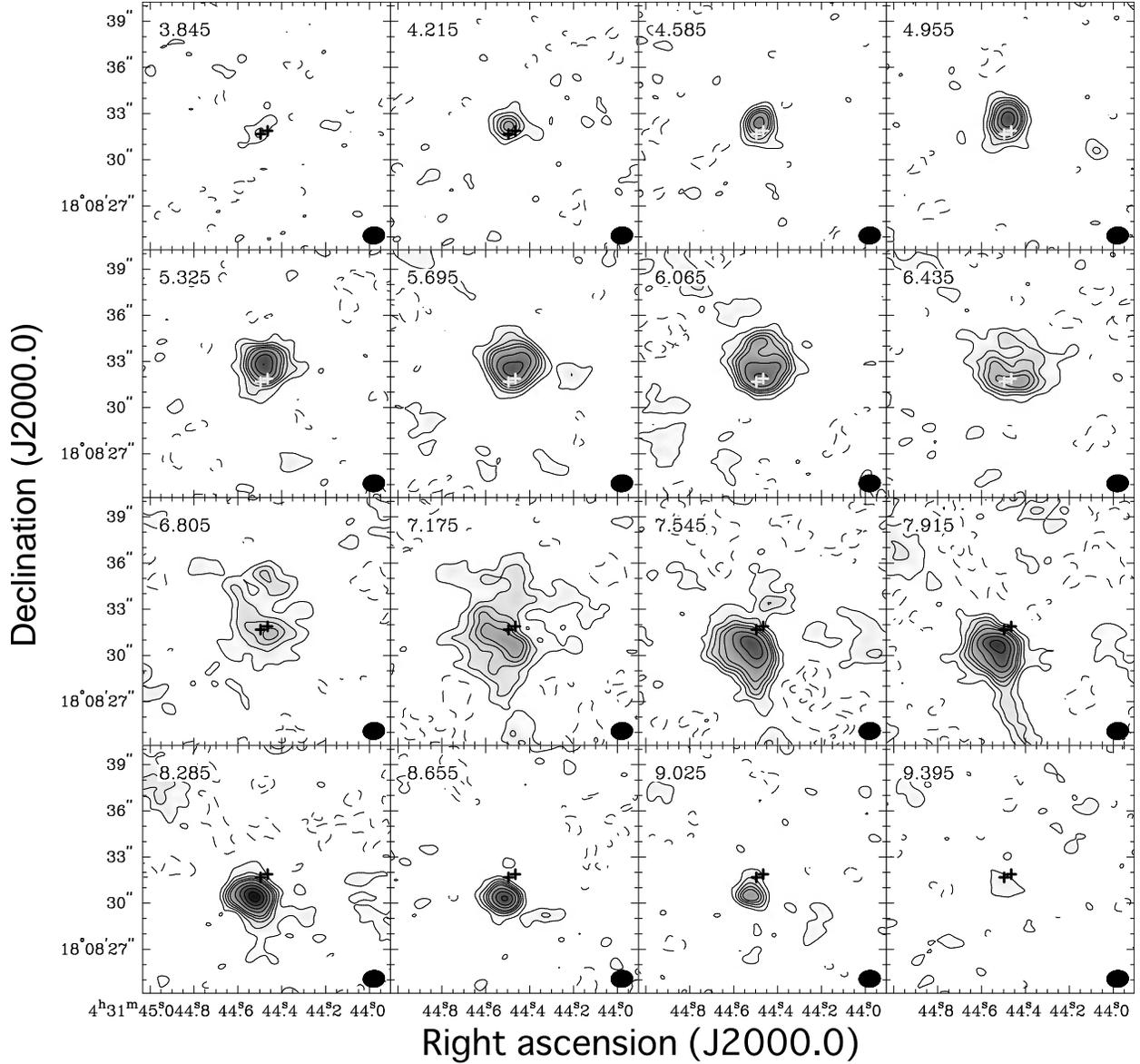}
\caption{Velocity-channel C$^{18}$O (3--2) maps made with the newly-acquired and previously published (Paper~I) data combined.  Contour levels are from 2$\sigma$ in steps of 2$\sigma$ until 10$\sigma$, and then in steps of 4$\sigma$
(1$\sigma$ = 0.081 Jy beam$^{-1}$).   The LSR velocity is indicated at the top-left corner of each panel.  Crosses indicate the positions of the binary components of L1551~NE as in Figure \ref{mom0}.  The filled ellipse at the bottom-right corner of each panel corresponds to the synthesized beam as plotted at FWHM (1\farcs36 $\times$ 1\farcs06 at P.A. = -86$\degr$). 
\label{ch18}}
\end{figure}

\begin{figure}
\epsscale{1.0}
\plotone{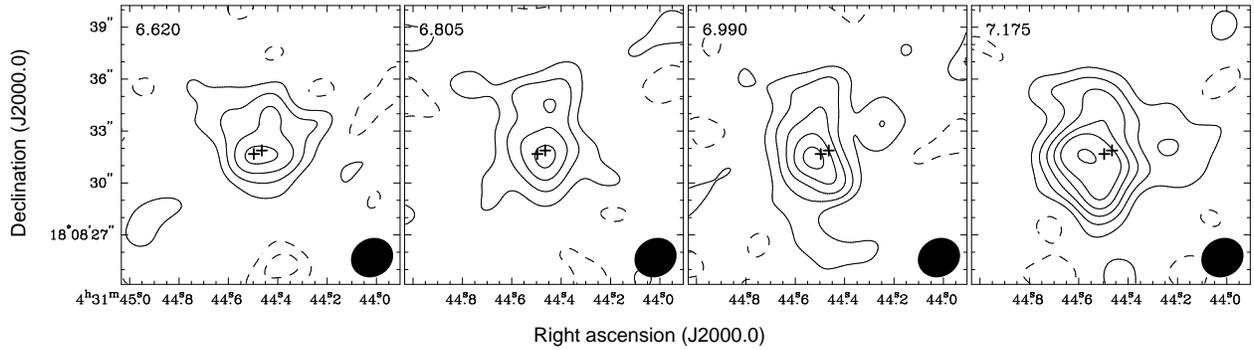}
\caption{Velocity-channel C$^{18}$O (3--2) maps made with the newly-acquired data only, which has a factor of 2 higher velocity resolution than the previously published data (Paper~I), at velocities close to the systemic velocity (6.9 km s$^{-1}$) .  Contour levels are from 2$\sigma$ in steps of 2$\sigma$ until 10$\sigma$, and then 14$\sigma$
(1$\sigma$ = 0.18 Jy beam$^{-1}$).  The LSR velocity is indicated at the top-left corner of each panel.  Crosses indicate the positions of the binary components of L1551~NE as in Figure \ref{mom0}.  The filled ellipse at the bottom-right corner of each panel corresponds to the synthesized beam as plotted at FWHM (2\farcs50 $\times$ 2\farcs20 at P.A. = -61$\degr$).
\label{chlv}}
\end{figure}

\begin{figure}
\epsscale{0.6}
\plotone{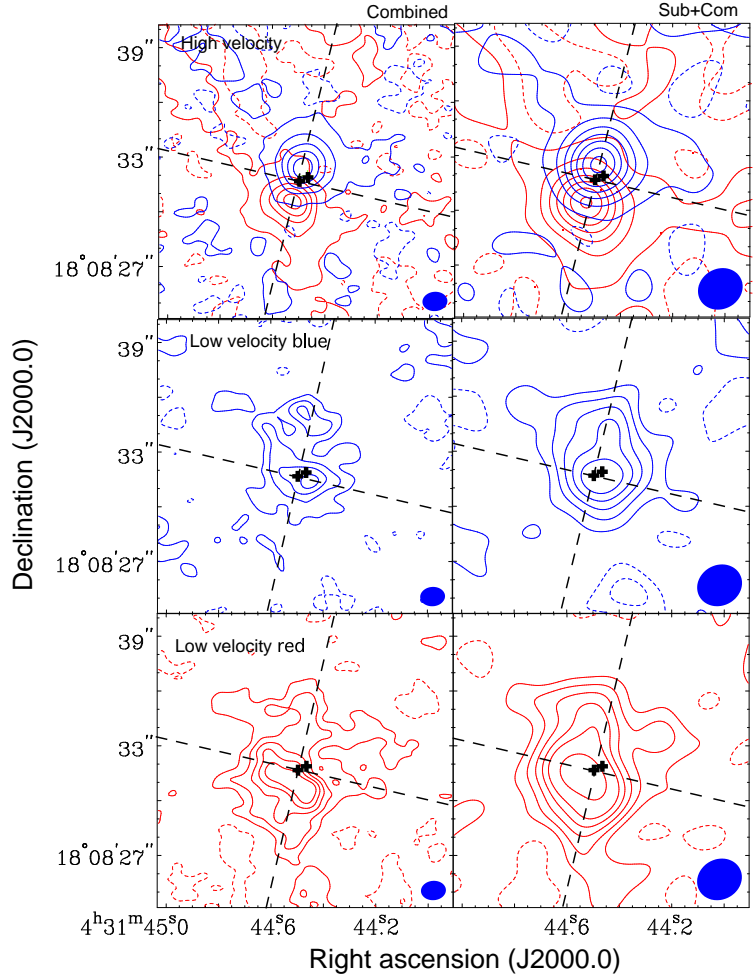}
\caption{Integrated C$^{18}$O (3--2) intensity maps over selected velocity ranges corresponding to the high-velocity (upper row), low-velocity blueshifted (middle row), and low-velocity redshifted (lower row) components (see text).  Left column shows images made with the newly-acquired and previous published data (Paper~I) combined, and right column just the newly-acquired data that provide higher sensitivity to large-scale structures.  Blueshifted emission is plotted in blue contours, and redshifted emission in red contours.  In the left panels, the (LSR) velocity range used to make the high-velocity blueshifted component spans 3.66--6.62 km s$^{-1}$, the high-velocity redshifted component 7.36--10.32 km s$^{-1}$, the low-velocity blueshifted component 6.62--6.99 km s$^{-1}$, and the low-velocity redshifted component 6.99--7.36 km s$^{-1}$.  In the right panels, the velocity range used to make the high-velocity blueshifted component spans 3.7525--6.5275 km s$^{-1}$, the high-velocity redshifted component 7.2675--10.0425 km s$^{-1}$, the low-velocity blueshifted component 6.5275--6.8975 km s$^{-1}$, and the low-velocity redshifted component 6.8975--7.2675 km s$^{-1}$.  Contour levels are plotted from 2$\sigma$ in steps of 8$\sigma$ for the high-velocity component (1$\sigma$ = 0.029 Jy beam$^{-1}$ and 0.0465 Jy beam$^{-1}$, respectively, in the left and right panels), and 2$\sigma$ in steps of 2$\sigma$ until 10$\sigma$ and then 14$\sigma$ for the low-velocity component (1$\sigma$ = 0.081 Jy beam$^{-1}$ and 0.127 Jy beam$^{-1}$, respectively, in the left and right panels).  Crosses and dashed lines are the same as in Figure \ref{mom0}.  The filled ellipse at the bottom-right corner of each panel corresponds to the synthesized beam as plotted at FWHM, with sizes as specified earlier in Figure \ref{ch18} for the left column and Figure \ref{chlv} for the right column.
\label{br18}}
\end{figure}

\begin{figure}
\epsscale{0.8}
\plotone{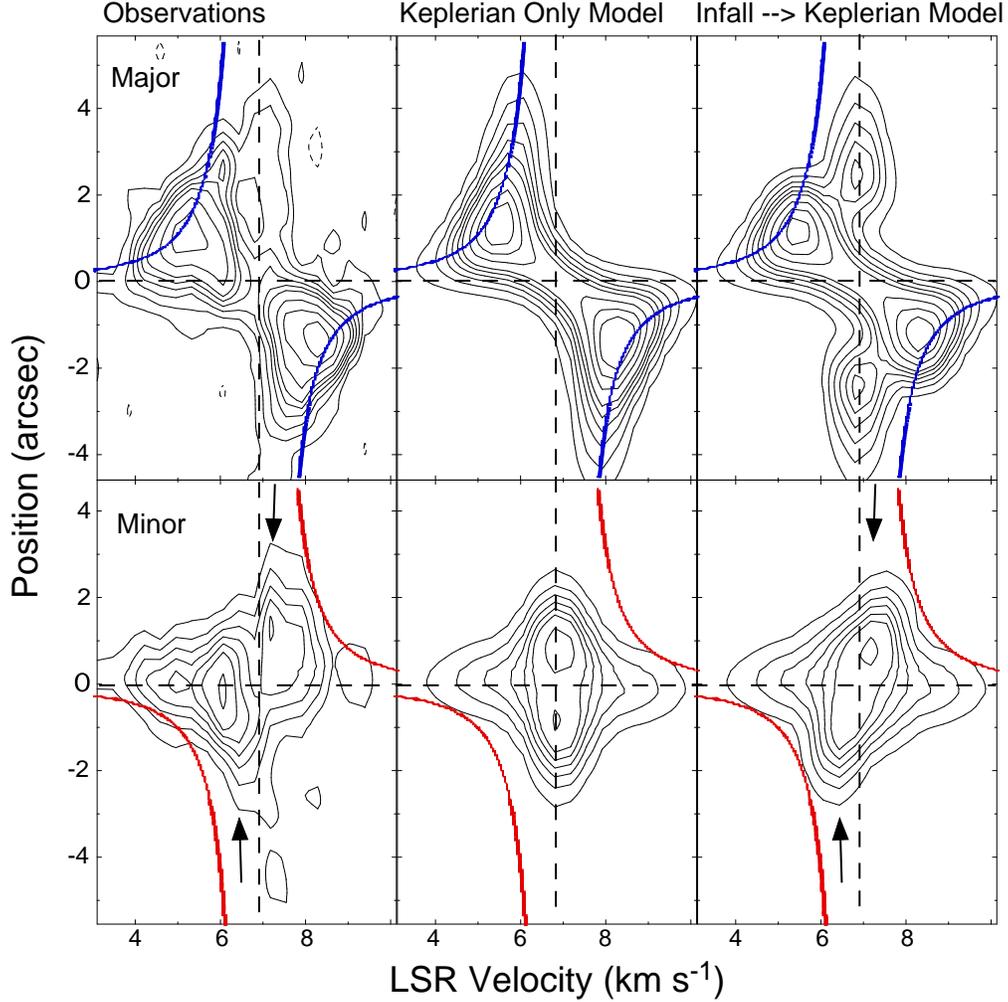}
\caption{Position-Velocity (P-V) diagrams along P.A. = 347$\degr$ in the upper row and P.A. = 77$\degr$ in the lower row and centered on Source A.  These position angles correspond to the major and minor axes, respectively, of the Keplerian circumbinary disk reported in Paper~I.  P-V diagrams in the left column are derived from the C$^{18}$O (3--2) maps made from the newly-acquired combined with the previously published (Paper~I) data.  P-V diagrams in the middle panels are derived from our model maps comprising a Keplerian circumbinary disk around a central stellar mass of 0.8 $M_{\odot}$, and observed at an inclination angle of 62$\degr$.  P-V diagrams in the right panel are from our model maps comprising a circumbinary disk exhibiting Keplerian motion out to a radius of 300~AU and infall beyond this radius at a constant velocity of $\sim$0.6 km s$^{-1}$.   Both models include an extended envelope component with properties as described in the text, and the resultant model images sampled with the $uv$-coverage attained in our combined observations with the SMA.  Contour levels are from 2$\sigma$ in steps of 2$\sigma$ until 12$\sigma$, and then in steps of 4$\sigma$ (1$\sigma$ = 0.081 Jy beam$^{-1}$). Vertical dashed lines indicate the systemic velocity for L1551~NE of 6.9 km s$^{-1}$.  Blue curves show the Keplerian rotation curve for a central stellar mass of 0.8 $M_{\odot}$ and observed at an inclination angle of 62$\degr$, and red curves free-fall in the plane of the Keplerian circumbinary disk onto a central stellar mass of 0.8 $M_{\odot}$.  Black arrows denote the velocity gradient detected along the minor axis.
\label{pv}}
\end{figure}

2. The velocity field of the circumbinary disk-like structure can be decomposed into two distinct components as seen in the velocity-channel maps of Figures~\ref{ch18}--\ref{br18} and most clearly in the corresponding P-V diagrams of Figure~\ref{pv}.  One component comprises high-velocity ($>$0.5 km s$^{-1}$) blueshifted and redshifted emissions located towards the north and south, respectively, of the binary protostellar system, as seen in Paper~I.  The other component comprises low-velocity ($<$0.5 km s$^{-1}$) blueshifted and redshifted emissions located towards the west and the east, respectively, of the binary protostellar system, seen here for the first time.  The high-velocity blueshifted and redshifted components are distinctly separated in space and aligned along the major axis of the circumbinary disk.  Their velocity field is consistent with Keplerian motion, and so these components are associated with the Keplerian circumbinary disk identified in Paper~I. {\bf Conversely}, the low-velocity components shows weak redshifted emission to the north in addition to strong redshifted emission to the south (the latter just like the high-velocity redshifted component) of the binary protostellar system, as well as an east (red) to west (blue) velocity
gradient along the minor axis of the circumbinary disk.
\\

We made a toy model comprising a circumbinary disk-like structure that exhibits pure Keplerian motion (for a binary protostellar mass of 0.8 $M_{\odot}$, as derived in Paper~I) in its inner region and pure infalling motion at a constant velocity in its outer region, and which is surrounded by a diffuse envelope.  From this toy model, we generated model maps for different Keplerian radii and infall velocities.  We then made mock observations of these model maps with the same $uv$-coverage as that attained in our observations with the SMA to generate simulated-observed model maps, and from the latter derived simulated-observed model P-V diagrams to compare with the observed P-V diagrams.  We found that:

3.  The inclusion or exclusion of a diffuse envelope, the parameters for which we could only roughly estimate, do not change in an important manner the simulated-observed model P-V diagrams.  The best-fit toy model requires motion in the disk to change from (pure) Keplerian to (pure) infall at a radius of $\sim$300~AU, the Keplerian radius.  The low-velocity blueshifted and redshifted components mentioned above trace primarily the outer infalling portion of the circumbinary disk.  The best-fit infall velocity is 0.6 km s$^{-1}$, which is smaller than the expected free-fall velocity of $\sim$2.2 km s$^{-1}$ at the Keplerian radius.  We tried, but were unable, to connect the motions within and beyond the Keplerian radius with a power-law rotational profile in a manner compatible with the observed P-V diagrams, suggesting that the transition between infall and rotation occurs over a region much narrower than can be studied in our observations.
\\

We considered the implications of our results for the formation of Keplerian circumbinary disks around protostellar systems:

4.  The lack of detectable rotation in the infalling component implies that angular momentum is effectively transported outwards in the outer region (i.e., beyond the Keplerian radius) of the circumbinary disk.  On the other hand, angular momentum is not effectively transported outwards in the inner region (i.e., within the Keplerian radius) of the same disk.  Borrowing in part ideas from Machida et al. (2011a; 2011b) but also adding our own, we suggest that this difference in effectiveness at which angular momentum is transported outwards may be caused by the very different plasma $\beta$, the ratio of the gas to magnetic pressure, along with the different ionization fractions of the molecular gas in these two regions.  Beyond the Keplerian radius, the plasma $\beta$ is relatively small and the ionization fraction relatively high so that magnetic fields have effective control of the molecular gas, allowing angular momentum to be efficiently transported outwards through magnetic breaking.  Within the Keplerian radius, the plasma $\beta$ increases whereas the ionization fraction decreases, so that magnetic fields no longer have effective control of the molecular gas and hence magnetic braking becomes ineffective.  Consistent with the idea of strong gas pressure near the Keplerian radius, the inferred infall velocity is smaller than the expected free-fall velocity near the Keplerian radius, presumably because the infalling material is decelerated as it moves into regions of higher gas pressure.  The transition between good and poor magnetic control of the molecular gas may occur quite abruptly not only at a certain radius, but also at a particular stage in the evolution of a protostellar system as the mass and hence density of the circumbinary disk grows.

5.  The inferred mass-accretion rate onto the Keplerian circumbinary disk is $\sim$9.6 $\times$ 10$^{-7}$ $M_{\odot}$ yr$^{-1}$.  Given that the mass in the surrounding dense condensation is $\sim$0.39 $M_{\odot}$, there is sufficient material to feed the Keplerian circumbinary disk at the abovementioned mass-accretion rate for the next $\sim$4.1 $\times$ 10$^{5}$ yr, which is comparable to the typical timescale estimated for the Class I evolutionary stage.  A large fraction of this material, however, must either accumulate onto the central protostellar system and/or be ejected or entrained in an outflow for the circumbinary disk to remain stable throughout its subsequent evolution.\\

%Our results of L1551 NE, as well as recent interferometric observations of Class I protostars, suggest that large-scale ($r$ $\gtrsim$100 AU) Keplerian disks must be formed in the embedded protostellar stage --> this conclusion is already in Paper 1.  Do not need to repeat here.

We close by offering a few thoughts on what future work ought to be conducted to better inform models to reproduce the Keplerian circumbinary disk of L1551~NE as well as test the ideas proposed above for the differences in effectiveness at which angular momentum is transported outwards within and beyond the Keplerian radius in the disk.  
%A region of immediate interest is the transition region between the outer predominantly infalling and inner predominantly centrifugally-supported disk.
Our observations with the SMA indicate that the transition from predominantly infalling to predominantly Keplerian motion occurs within an annular strip having an angular width of $\sim$2$\arcsec$ where
the observed radial velocity changes by $\sim$0.5 km s$^{-1}$ (Figure \ref{pv}).  Observations with ALMA should be able to easily probe the kinematics of this transition region, and reveal how the infall and rotational velocities change with radius so as to compare with theoretical models (e.g., such as those by Machida et al. 2011a; 2011b).  Single-dish mapping of L1551 NE in C$^{18}$O (3--2), combined with our SMA data, will permit studies of infall and rotation, and hence the transport of angular momentum, from the outer region of the envelope (at $\gtrsim$1500 AU) to the Keplerian circumbinary disk.  Observation of ionic molecules such as HCO$^{+}$ and N$_{2}$H$^{+}$ at high angular resolutions can be used to look for differences in ionization \citep[e.g.,][]{ben98,cas98,wil98} between the Keplerian circumbinary disk and the surrounding infalling regions, and hence test our idea that differences in ionization fractions play an important role in how effectively magnetic fields can exert control over molecular gas in the different regions.
%These measurements should reveal the relative difference of the magnetic effects on the gas dynamics between the infalling and disk components, and verify our hypothesis that difference of the magnetic braking causes difference of the efficiency of the angular momentum transfer between the outer infalling and the inner Keplerian region.

\acknowledgments
We are grateful to P. T. P. Ho and N. Ohashi for their fruitful discussions. We would like to thank all the SMA staff supporting this work.  S.T. acknowledges a grant from the National Science
Council of Taiwan (NSC 99-2112-M-001-013-MY3) in support of this work.  J. L. is supported by the GRF Grants of the Government of the Hong Kong SAR under HKU 703512P for conducting this research.

\clearpage
%%%%%%%%%%%
% Figures %
%%%%%%%%%%%

% Combined High-vel Blue 20 (3.845) - 27 (6.435)   --> 3.66 - 6.62 km s-1
%                                  Red 30 (7.545) - 37 (10.135) -> 7.36 - 10.32 km s-1
%  --> Contour 2sigma in steps of 8sigma (1sigma=0.029 Jy beam-1)
% Low-vel Blue 28 (6.805) --> 6.62 - 6.99 km s-1
% Low-vel Red 29 (7.175)  --> 6.99 - 7.36 km s-1
% --> Contour 2sigma in steps of 2sigma until 10sigma, then 14 sigma (1sigma=0.081 Jy beam-1)
%
% 0.185 / 2 = 0.0925
% (Sub)Comp Low-vel Blue 53 (6.620) - 54 (6.805) --> 6.5275 - 6.8975 km s-1
%                     Low-vel Red 55 (6.990) - 56 (7.175) --> 6.8975 - 7.2675 km s-1
% Contour 2sigma in steps of 2sigma until 10sigma, then 14 sigma (1sigma=0.127 Jy beam-1)
%
% (sub)Comp High Blue 38 (3.845) - 52 --> 3.7525 - 6.5275 km s-1
%                             Red 57 - 71 (9.95)    --> 7.2675 - 10.0425 km s-1
% --> Contour 2sigma in steps of 8sigma (1sigma=0.0465 Jy beam-1)

% \begin{figure}
% \epsscale{0.8}
% \plotone{PVobs_models3noenv.eps}
% Same as Figure \ref{pv} but for the models without the extended envelope component.
% Contour levels are from 2$\sigma$ in steps of 2$\sigma$ until 12$\sigma$,
% and then in steps of 4$\sigma$ (1$\sigma$ = 0.078 Jy beam$^{-1}$).
% \caption{\label{pvno}}
% \end{figure}

%%%%%%%%%%%
% Table 1 %
%%%%%%%%%%% 
\clearpage
\begin{deluxetable}{lcc}
\tabletypesize{\scriptsize}
\tablecaption{Parameters for the SMA Observations of L1551 NE \label{tbl-1}}
\tablewidth{0pt}
% \tablewidth{250pt} % use this in emulateapj
\tablehead{\colhead{Parameter} & \multicolumn{2}{c}{Value}\\
\cline{2-3}
\colhead{} & \colhead{2011 December 28} & \colhead{2012 January 14} }
\startdata
Number of Antennas &8 &7  \\
Configuration & Compact &Subcompact \\
Flux Calibrator &Uranus &Callisto\\
Gain Calibrator & \multicolumn{2}{c}{0423-013, 0530+135}\\
Flux (0423-013 Upper) &2.09 Jy &1.99 Jy\\
Flux (0423-013 Lower) &2.08 Jy &1.99 Jy\\
Flux (0530+135 Upper) &0.96 Jy &1.02 Jy\\
Flux (0530+135 Lower) &0.95 Jy &1.08 Jy\\
Passband Calibrator &3c84, 3c279, Uranus &bllac, 3c279, Callisto, Mars \\
System Temperature (DSB) &$\sim$250 - 1000 K &$\sim$500 - 1500 K\\
Absolute Flux Uncertainty &\multicolumn{2}{c}{$\sim$30$\%$}\\
\enddata
\end{deluxetable}

\begin{deluxetable}{lcc}
\tabletypesize{\scriptsize}
\tablecaption{Parameters of the C$^{18}$O (3--2) image cube taken with the SMA subcompact, compact, and
the extended configurations \label{tbl-2}}
\tablewidth{0pt}
% \tablewidth{250pt} % use this in emulateapj
\tablehead{\colhead{Parameter} &\colhead{All the Config.} &\colhead{Sub. + Compact Config.}}
\startdata
($\alpha_{J2000.0}$, $\delta_{J2000.0}$) & \multicolumn{2}{c}{(04$^{\rm h}$ 31$^{\rm m}$ 44$^{\rm s}$.47, 18$^{\circ}$ 08$\arcmin$ 32\farcs2)} \\
Primary Beam HPBW& \multicolumn{2}{c}{$\sim$37$\arcsec$}\\
Synthesized Beam HPBW &1\farcs36 $\times$ 1\farcs06 (P.A. = -86$\degr$) &2\farcs50 $\times$ 2\farcs20 (P.A. = -61$\degr$)\\
Frequency Resolution & 406.25 (kHz) $\sim$0.37 (km s$^{-1}$) &203.125 (kHz) $\sim$0.185 (km s$^{-1}$) \\
Baseline Coverage &6.2 - 249.0 (k$\lambda$) &6.2 - 84.5 (k$\lambda$)\\
Conversion Factor & 7.82 (K / J beam$^{-1}$) &2.05 (K / J beam$^{-1}$) \\
rms noise level &0.078 (Jy beam$^{-1}$) &0.18 (Jy beam$^{-1}$)\\
\enddata
\end{deluxetable}

\end{document}